\documentclass{elsarticle}
\usepackage{xcolor}
\usepackage{bm}
\usepackage{graphicx}
\usepackage{amsfonts}
\usepackage{amsmath}
\usepackage{subfigure}
\usepackage{float}
\usepackage{lineno,hyperref}
\modulolinenumbers[5]

\begin{document}

\begin{frontmatter}

\title{Personality Traits and Echo Chambers on Facebook}

\author{Alessandro Bessi}
\corref{mycorrespondingauthor}
\cortext[mycorrespondingauthor]{Corresponding author}
\ead{alessandro.bessi@iusspavia.it}

\address[mymainaddress]{IUSS Institute for Advanced Study, Pavia, Italy}
\address[mysecondaryaddress]{IMT Institute for Advanced Studies, Lucca, Italy}

\begin{abstract}
In online social networks, users tend to select information that adhere to their system of beliefs and to form polarized groups of like minded people. Polarization as well as its effects on online social interactions have been extensively investigated. Still, the relation between group formation and personality traits remains unclear.  A better understanding of the cognitive and psychological determinants of online social dynamics might help to design more efficient communication strategies and to challenge the digital misinformation threat.
In this work, we focus on users commenting posts published by US Facebook pages supporting scientific and conspiracy-like narratives, and we classify the personality traits of those users according to their online behavior. We show that different and conflicting communities are populated by users showing similar psychological profiles, and that the dominant personality model is the same in both scientific and conspiracy echo chambers. Moreover, we observe that the permanence within echo chambers slightly shapes users' psychological profiles. Our results suggest that the presence of specific personality traits in individuals lead to their considerable involvement in supporting narratives inside virtual echo chambers.
\end{abstract}

\end{frontmatter}

%\linenumbers

\section{Introduction}
In online social media, users show the tendency to select information that confirms their preexisting beliefs. Being influenced by confirmation bias and selective exposure, they join virtual echo chambers --- i.e. polarized communities populated by like-minded users. Polarization as well as its effects on online social dynamics have been extensively investigated \cite{bakshy2015exposure,conover2011political,adamic2005political,an2013fragmented,mocanu2015collective,bessi2015science,bessi2014economy,sunstein2002law}. In particular,  discussion within like-minded people seems to influence negatively users emotions and to enforce group polarization \cite{zollo2015emotional}. Moreover, experimental evidence shows that confirmatory information gets accepted even if containing deliberately false claims \cite{bessi2015science}, while dissenting information are mainly ignored or might even increase group polarization \cite{zollo2015debunking}. Furthermore, recent studies clearly show that confirmation bias, more than algorithms of content promotion \cite{bessi2016users}, plays a pivotal role in the formation of echo chambers \cite{del2016spreading}. Finally, users on social media aim at maximizing the number of likes, and often information, concepts and debate get flattened and oversimplified \cite{dewey2012public,habermas2015between}.

The cognitive and psychological dimensions of users either as individuals or as a part of a group influence and shape online social interactions. Indeed, a large research effort has been payed in studying the interplay between personality of users and their online behavior \cite{amichai2010social,golbeck2011predicting,oberlander2006whose,quercia2011our,kosinski2013private,muscanell2012make,marriott2014true,worth2014personality,michikyan2014can,kern2014online}.  Still, the relation between group formation and personality traits remains unclear. 

Psychologists describe personality along five dimensions known as the \emph{Big Five} \cite{goldberg1992development,norman1963toward}. According to this framework, such five dimensions contain most known personality traits and represent the basic structure behind all personalities \cite{o2002quantitative}. In particular, these dimensions are extraversion, emotional stability, agreeableness, conscientiousness, and openness. Extraversion is defined as the state of being concerned primarily with things outside the self. Emotional Stability refers to an individual's ability to remain calm when faced with pressure or stress. Agreeableness reflects a tendency to be compassionate and cooperative rather than suspicious and antagonistic towards others. Conscientiousness is a tendency to show self-discipline and act dutifully. Finally, Openness is related to curiosity and to a general appreciation for unusual ideas, imagination, and novel experiences.

In this paper, we aim to understand the personality traits driving the adoption of a specific narrative and the emergence of echo chambers. By means of a well established unsupervised personality recognition approach \cite{celli2012unsupervised}, we want to understand whether users in echo chambers have similar personality traits, and whether a specific narrative attracts certain psychological profiles. 

In particular, we focus on users commenting posts published by US Facebook pages supporting the scientific narrative (Science) and the conspiracy-like one (Conspiracy). We choose to consider these specific narratives for two main reasons: a) Science and Conspiracy are two very distinct and conflicting narratives; b) scientific pages share the main mission to diffuse scientific knowledge, whereas conspiracy-like pages diffuse myth narratives, hoaxes, false news, and controversial information. Thus, our contribution is twofold. First, we provide a statistical characterization of the personality traits of users embedded in conflicting echo chambers. Moreover, we provide additional insights that might be crucial to develop strategies to mitigate the spreading of misinformation online. Indeed, the World Economic Forum listed massive digital misinformation as one of the main threats for the modern society \cite{howell2013digital,quattrociocchi2016misinformation} and, despite different debunking strategies have been proposed, unsubstantiated rumors and false news keep proliferating in polarized communities emerging in online social networks \cite{bessi2014social, bessi2015viral,zollo2015emotional,zollo2015debunking,bessi2015trend}.

In this work, we perform a comparative analysis on personality traits of users engaged with different and conflicting narratives. 
We measure extraversion, emotional stability, agreeableness, conscientiousness, and openness of about $30K$ users who made more than $3M$ comments in a time span of $5$ years (Jan 2010 --- Dec 2014).

We find that such personality traits are similarly distributed within the polarized communities, with the exception of the emotional stability, which is higher in users supporting the conspiracy-like narrative. Moreover, we find very similar and significant correlations between personality traits within different echo chambers. Furthermore, we show that the prevalent personality model is the same in both the observed echo chambers. In particular, the most common supporters of Science and Conspiracy tend to enjoy interactions with close friends (low extraversion), are emotionally stable (high emotional stability), are suspicious and antagonistic towards others (low agreeableness), engage in antisocial behavior (low conscientiousness), and have unconventional interests (high openness). Finally, we observe very weak Pearson's correlations between the number of comments made by users and their personality traits.  Such a result provides meaningful insights towards the relationship between the psychological profile of users and their commitment inside polarized online communities. Indeed, the weak correlations between users' activity and their personality traits indicate that the permanence within echo chambers slightly shapes users' psychological profiles. Rather, our analysis suggests that the presence of specific personality traits in individuals lead to their considerable involvement in supporting narratives inside virtual echo chambers.

\section{Methods}

\subsection{Dataset}
We analyze users commenting on $413$ US public Facebook pages supporting conflicting narratives --- i.e. Science and Conspiracy --- within a temporal window of $5$ years (Jan 2010 to Dec 2014). Science pages aim at diffusing scientific knowledge and rational thinking, whereas Conspiracy pages diffuse controversial information, usually lacking supporting evidence and most often contradictory of the official news. Such a space of investigation is defined with the same approach as in \cite{del2016spreading, bessi2015science}, with the support of different Facebook groups very active in monitoring conspiracy narratives. 

On Facebook, a \emph{like} stands for a positive feedback to the post, whereas a \emph{comment} is the way in which users express their personality and online collective debates take form. 

Here, we consider a user as embedded in the Science (Conspiracy) echo chamber if she is polarized towards Science (Conspiracy) --- i.e. if and only if she has more than the $95\%$ of their total likes on posts published by Science or Conspiracy pages. Moreover, we analyze only users who left at least $50$ comments in order to provide reliable estimates of the personality traits. The final dataset is composed by $25,767$ users supporting Science who left $2,620,733$ comments, and $6,262$ users supporting Conspiracy who left $666,592$ comments.

The entire data collection process has been carried out exclusively through the Facebook Graph API, which is publicly available. We used only public available data. The pages from which we downloaded data are public Facebook entities.

\subsection{Personality Model Recognition}
In this work, we represent the Big Five dimensions \cite{goldberg1992development,norman1963toward} --- i.e. extraversion, emotional stability, agreeableness, conscientiousness, and openness --- as discrete numerical variables that can take both positive and negative values. For each dimension, a positive value indicates the presence of the personality trait; a negative value indicates the presence of the reversed personality trait; a value equal to zero indicates a balance between the two extremes of the spectrum. For instance, if we consider the extraversion, a positive value reflects an extrovert individual; a negative value reflects an introvert individual; a value equal to zero reflects an ambivert individual.

To assign a personality model to each user, we rely on an established unsupervised personality recognition approach \cite{celli2012unsupervised} which leverages a series of statistically significant correlations between linguistic features and personality traits \cite{mairesse2007using} --- e.g. extraversion is positively correlated with the use of first person singular pronouns and negatively correlated with the use of parentheses, while emotional stability is negatively correlated with the use of exclamation marks and positively correlated with the use of words longer than six letters.

The classification strategy may be summarized as follows.
In the first step, for each user we analyze her comments and compute the mean count for the following features:
\begin{enumerate}
	\item \textbf{ap}: all punctuation;
	\item \textbf{cm}: commas;
	\item \textbf{em}: exclamation marks;
	\item \textbf{el}: external links;
	\item \textbf{im}: first person singular pronouns;
	\item \textbf{np}: negative particles;
	\item \textbf{ne}: negative emoticons;
	\item \textbf{nb}: numbers;
	\item \textbf{pa}: parenthesis;
	\item \textbf{pe}: positive emoticons;
	\item \textbf{pp}: prepositions;
	\item \textbf{qm}: question marks;
	\item \textbf{sl}: words longer than 6 letters;
	\item \textbf{sr}: first person (singular and plural) pronouns;	
	\item \textbf{sw}: vulgar words and expressions;
	\item \textbf{wc}: words;
	\item \textbf{we}: first person plural pronouns;
	\item \textbf{yu}: second person singular pronouns.	
\end{enumerate} 
In the second step, we compute the average values of the aforementioned features in the entire dataset. In the third step, we build a personality model for each user applying the following rule: if a user shows a feature correlating positively (negatively) with one personality trait and the value of that feature is greater than the average value of that feature, then the score of that personality trait is increased (decreased). Then, numerical values are turned into labels --- i.e. ``y", ``n", ``o" --- by checking if a value is positive, negative, or equal to zero. 

Finally, each user is represented by a personality model of five labels indicating, for each of the five dimensions, whether he has a given personality trait (``y") or its reversed (``n") or none of the two (``o"). For instance, a user represented by the personality model ``nyyoo" is an introvert, emotionally stable, agreeable individual.

\section{Results and Discussion}
In this work, we provide a statistical characterization of the personality traits of Facebook users embedded in conflicting echo chambers. In the next sections, we first compare the statistical distributions of personality traits of users supporting different narratives. Then, we analyze the correlations between such personality traits. Finally, we look for the prevalent personality models in the observed echo chambers.

\subsection{Distribution of Personality Traits}
\begin{figure}[H]
	\centering
	\subfigure[Extraversion]
	{\includegraphics[width=0.45\textwidth]{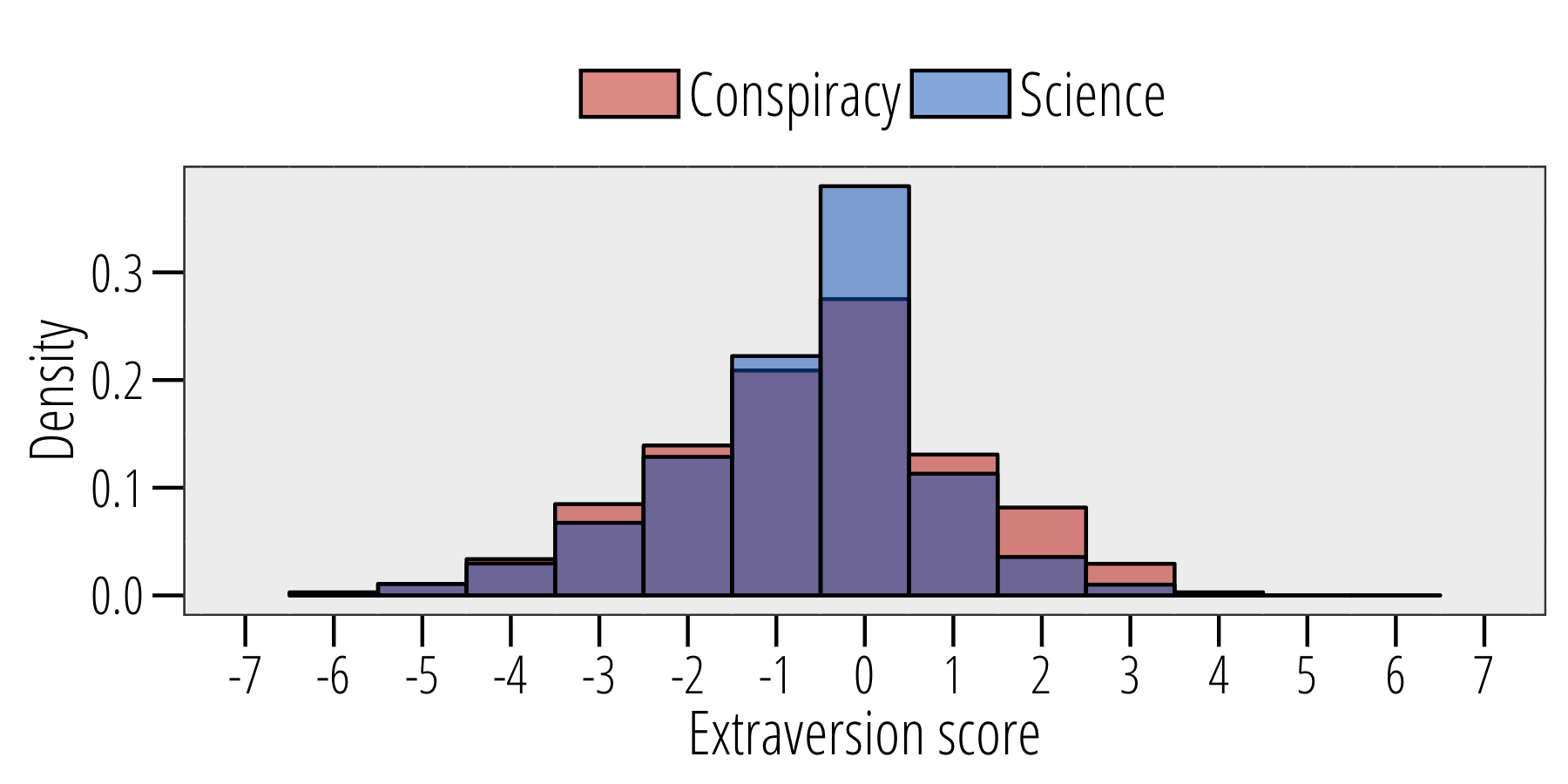} \label{fig:e}} 
	\subfigure[Emotional Stability]
	{\includegraphics[width=0.45\textwidth]{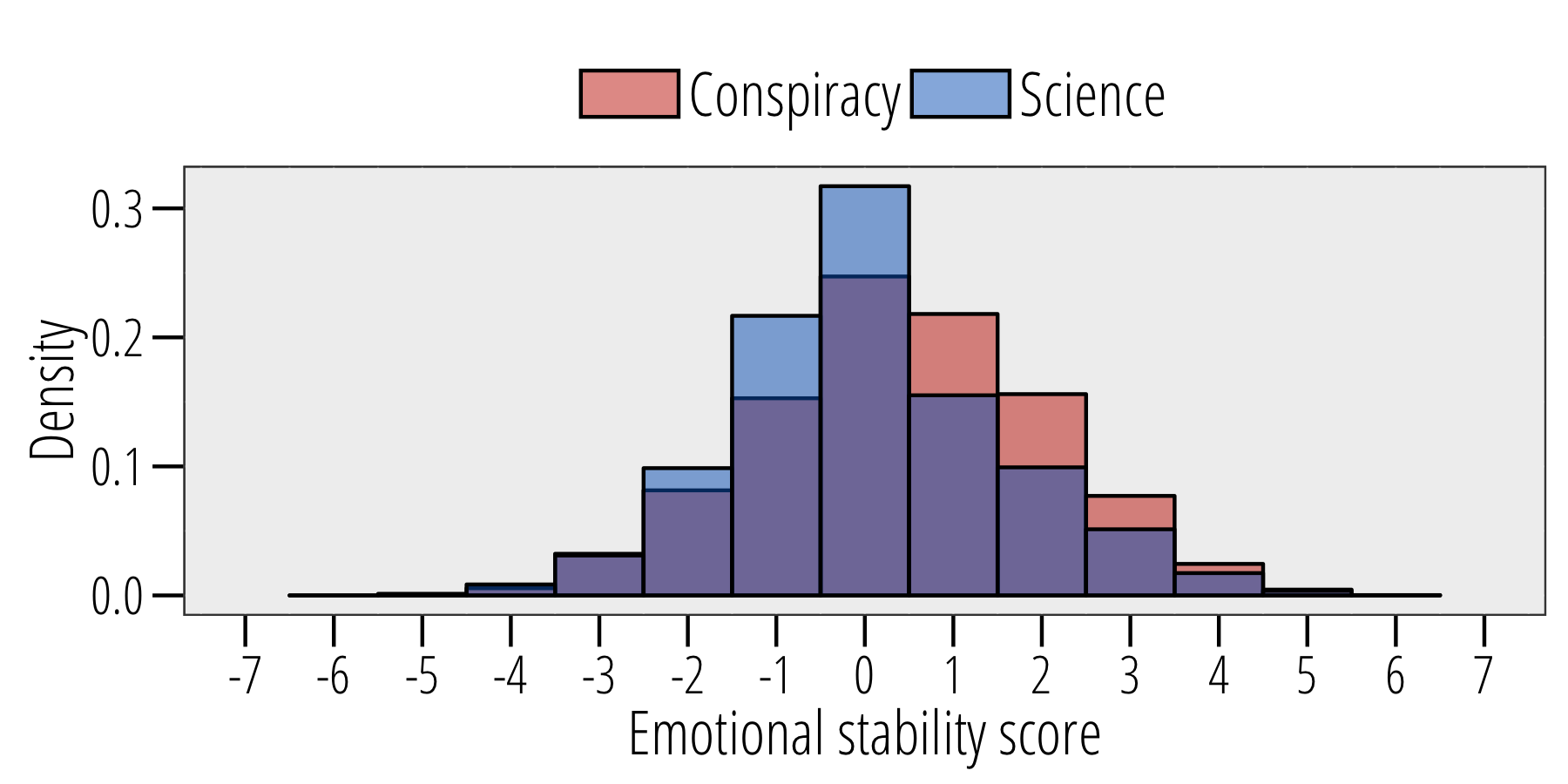} \label{fig:s}} 
	\subfigure[Agreeableness]
	{\includegraphics[width=0.45\textwidth]{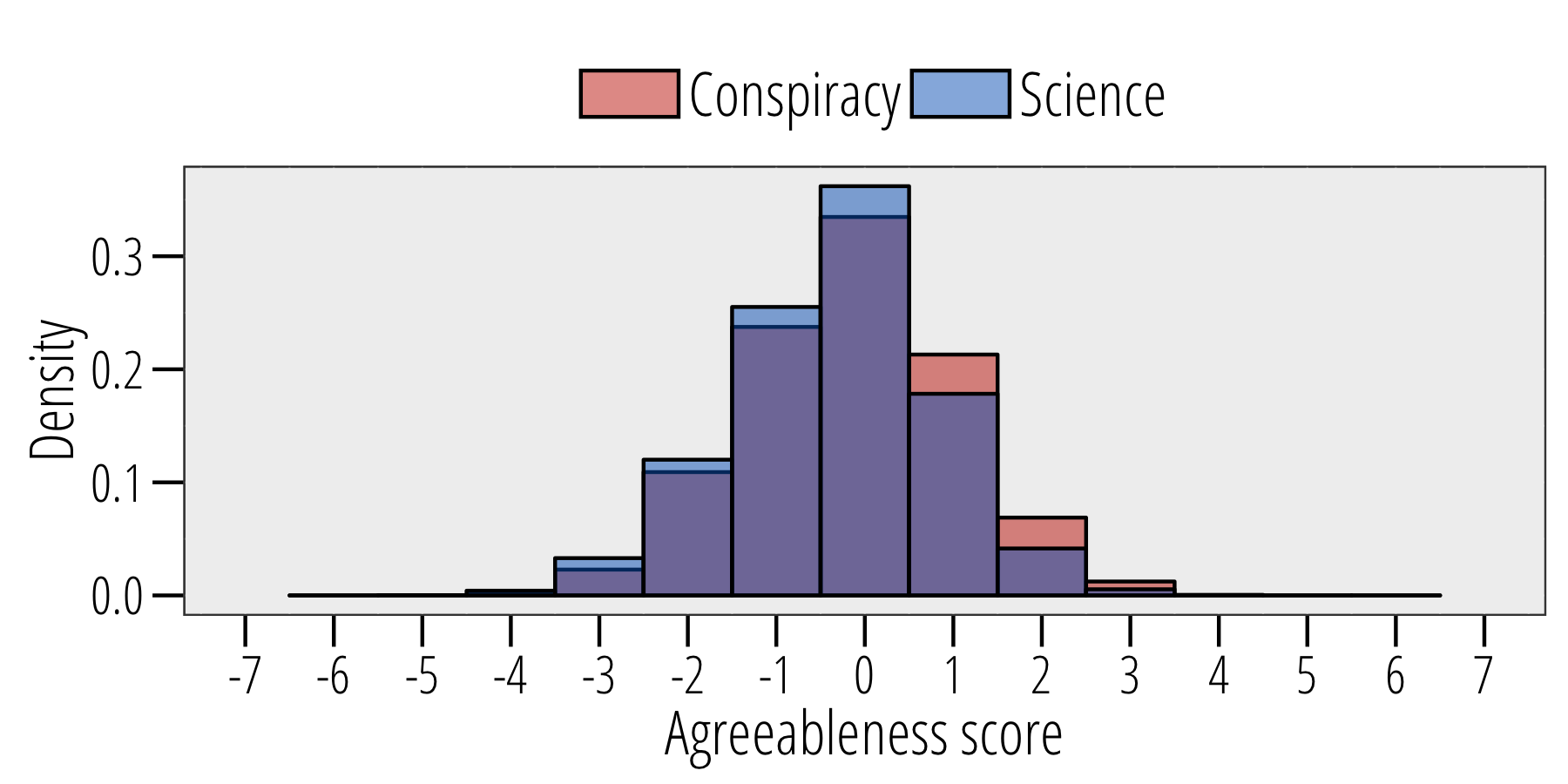} \label{fig:a}} 
	\subfigure[Conscientiousness]
	{\includegraphics[width=0.45\textwidth]{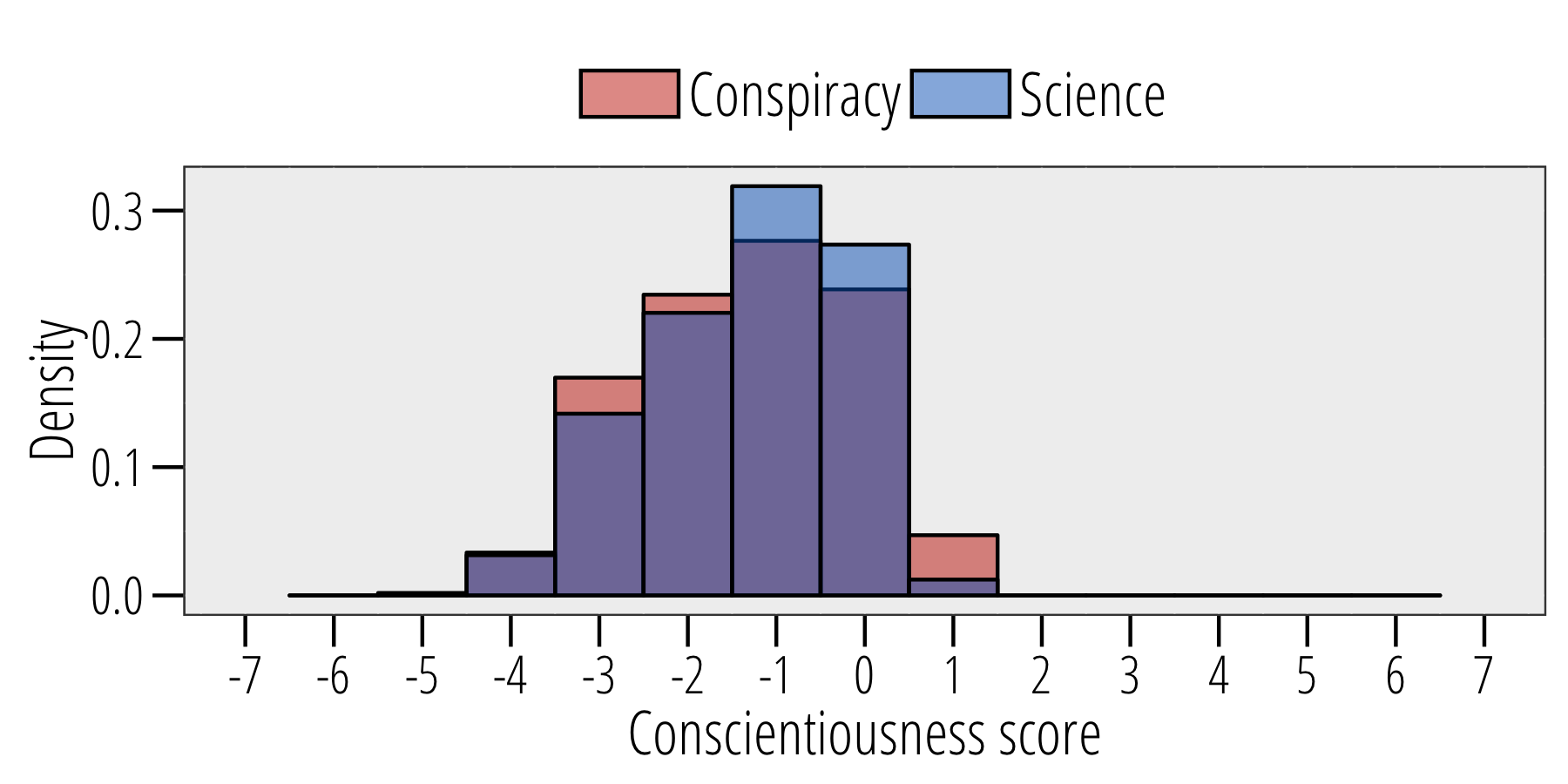} \label{fig:c}} 
	\subfigure[Openness]
	{\includegraphics[width=0.45\textwidth]{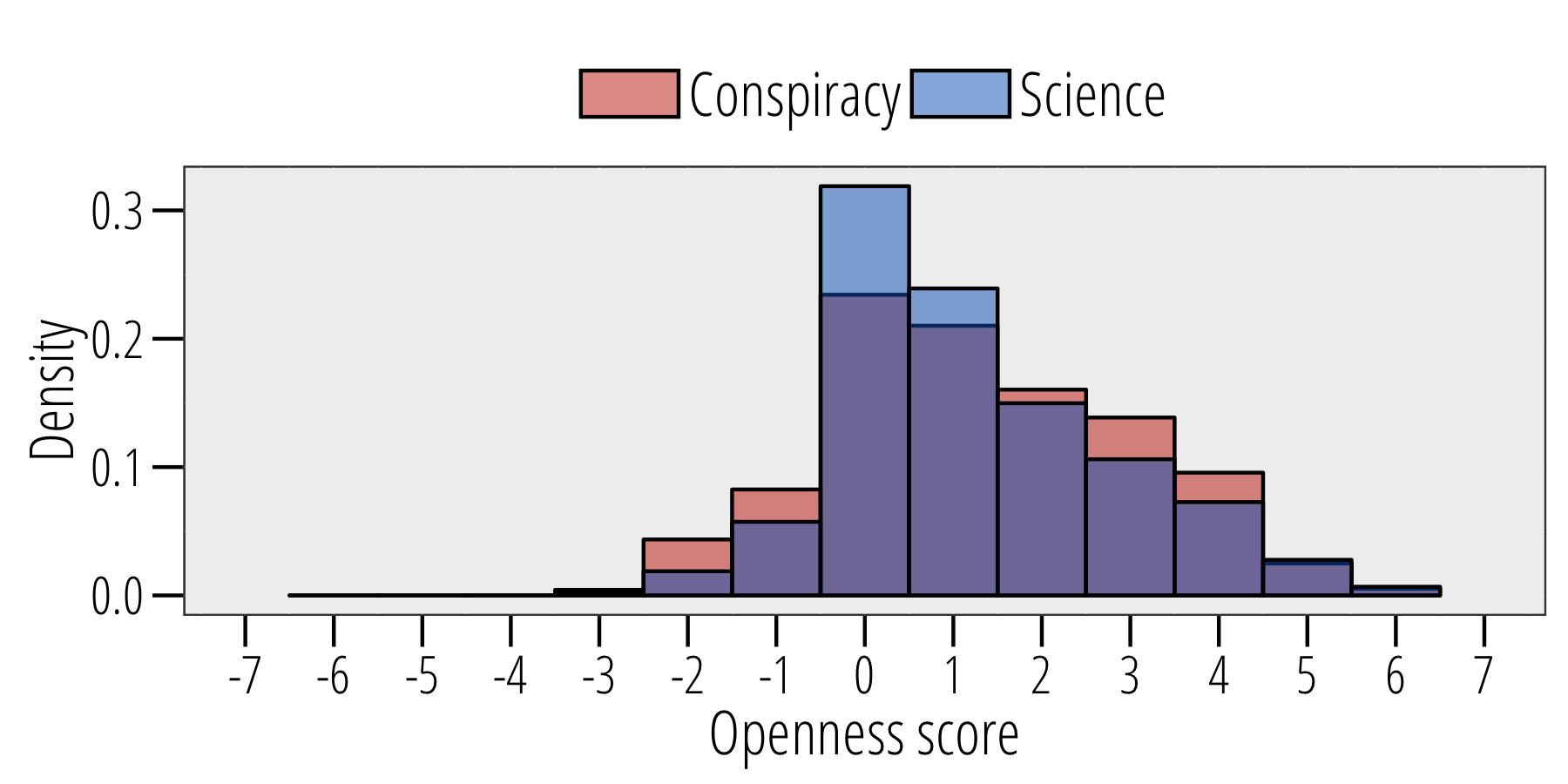} \label{fig:o}} 
	\caption{\textbf{Distribution of personality traits.} Statistical distributions of personality traits in different and conflicting echo chambers.}
	\label{fig:ptd_all}
\end{figure}

As a first step, we compute the statistical distributions of the five dimensions of personality for users embedded in conflicting echo chambers --- i.e. Science and Conspiracy. Figure \ref{fig:e} shows the distributions of Extraversion scores. Extraversion is defined as the state of being concerned primarily with things outside the self.  For Science supporters we find a mean score equal to $-0.65 (\pm 1.45)$, whereas for Conspiracy supporters the mean score is $-0.55 (\pm 1.70)$. In both echo chambers, the average extraversion scores indicate that users supporting Science and Conspiracy are slightly introvert. Introvert individuals are likely to enjoy time spent alone and find less reward in time spent with large groups of people, though they may enjoy interactions with close friends \cite{laney2002introvert}. 

Figure \ref{fig:s} shows the distributions of Emotional Stability scores. Emotional Stability refers to an individual's ability to remain calm when faced with pressure or stress. The mean score for Science supporters is $0.05 (\pm 1.56)$, whereas for Conspiracy supporters we find a mean score equal to $0.45 (\pm 1.65)$. Such results indicate a statistically significant (Mann-Whitney test, p-value $< 10^{-6}$) higher emotional stability in users supporting conspiracy narratives. Those who score low in emotional stability are emotionally reactive and vulnerable to stress. They are more likely to interpret ordinary situations as threatening, and minor frustrations as hopelessly difficult \cite{holt2012psychology}.

Figure \ref{fig:a} shows the distributions of Agreeableness scores. Agreeableness reflects a tendency to be compassionate and cooperative rather than suspicious and antagonistic towards others \cite{hogan1997handbook}. We find two similar distributions, with low levels of agreeableness in both echo chambers. In particular, the mean score for Science supporters is $-0.33 (\pm 1.18)$ and the mean score for Conspiracy supporters is $-0.14 (\pm 1.22)$. In both echo chambers we find a tendency to be suspicious and antagonistic towards others, especially for users supporting Science.

Figure \ref{fig:c} illustrates the distributions of Conscientiousness scores. Conscientiousness is a tendency to show self-discipline and act dutifully. People who score low on conscientiousness are more likely to engage in antisocial behavior \cite{ozer2006personality}. In both echo chambers we find low levels of conscientiousness, indicating low self-discipline as a specific personality trait of users embedded inside echo chambers supporting conflicting narratives. In particular, the average score for Science supporters is $-1.31 (\pm 1.15)$, whereas for Conspiracy supporters the mean score is $-1.34 (\pm 1.24)$.  Such results indicate the inclination to engage in antisocial behavior for both users supporting Science and Conspiracy.

Finally, Figure \ref{fig:o} illustrates the distributions of Openness scores.  Openness is related to curiosity and to a general appreciation for unusual ideas, imagination, and novel experiences \cite{mccrae1987validation}. The average score for Science supporters is $1.23 (\pm 1.60)$, whereas for Conspiracy supporters the mean score is $1.31 (\pm 1.75)$. In both echo chambers, we find positive levels of openness, indicating a tendency to have unconventional interests and a preference for the complex and ambiguous over the plain and the straightforward.

\begin{figure}[H]
	\includegraphics[width = \textwidth]{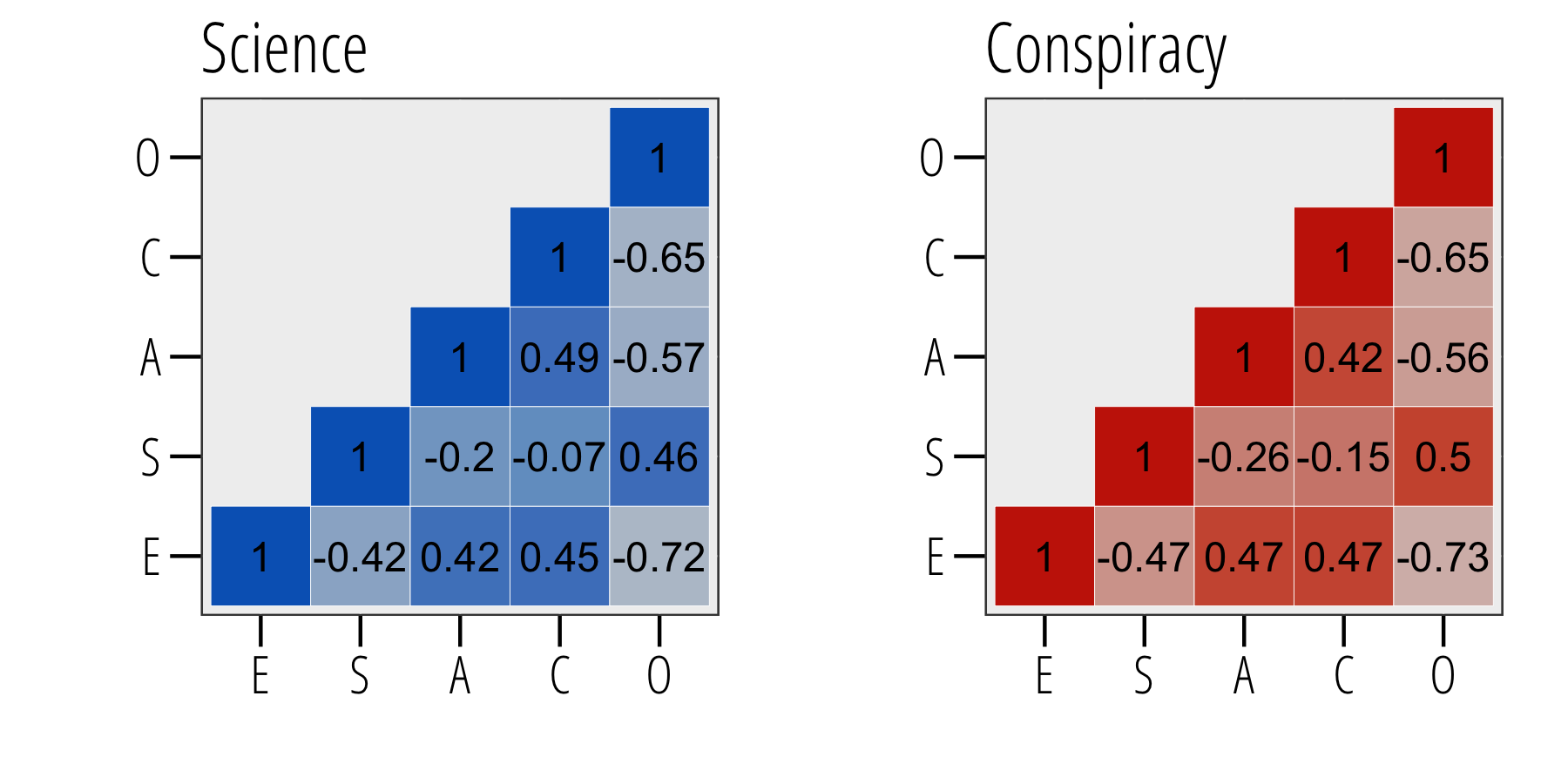}
	\caption{\textbf{Correlation matrices.} Correlation matrices of personality traits --- i.e. extraversion (E), emotional stability (S), agreeableness (A), conscientiousness (C), openness (O) --- of users supporting Science (left) and Conspiracy (right). We find a statistically significant similarity between the two matrices.}
	\label{fig:corr}
\end{figure}

To provide a better characterization of the environment under analysis, we study how different personality traits correlate within the two echo chambers. Figure \ref{fig:corr} shows the Pearson's correlation coefficients between the five personality traits of Science and Conspiracy supporters. By means of the Mantel test, we find a statistically significant (simulated p-value $< 0.01$, based on $10^{4}$ Monte Carlo replicates), high, and positive ($r = 0.996$) correlation between the correlation matrices of Science and Conspiracy supporters. 

Our analysis shows that conflicting narratives aggregate users with very similar personality traits. Users consume information according to their preferences, influenced by confirmation bias and selective exposure. However, the distributions of psychological traits within the two echo chambers are similar. In particular, users embedded in different echo chambers and supporting conflicting narratives tend to enjoy interactions with close friends (low extraversion), to be suspicious and antagonistic towards others (low agreeableness), to engage in antisocial behavior (low conscientiousness), and to have unconventional interests (high openness). Moreover, we assess that personality traits correlate in a statistically significant similar way within the two echo chambers.

\subsection{Personality and Echo Chambers}
As a further step, we want to identify the prevalent Personality Models (PM) inside the Science and Conspiracy echo chambers. Table \ref{tab:model} shows the top ten personality models of users supporting Science and Conspiracy. A personality model is characterized by five labels --- one for each of the \emph{Big Five} dimensions, i.e. extraversion, emotional stability, agreeableness, conscientiousness, openness --- indicating whether a user has a given personality trait (``y") or its reversed (``n") or none of the two (``o"). For instance, the personality model ``nyyoo" depicts users that are introvert, emotionally stable, and agreeable.

Our results show that, in both echo chambers, the dominant personality model is ``nynny", pointing out the strong prevalence of individuals that enjoy interactions with close friends (low extraversion), are emotionally stable (high emotional stability), suspicious and antagonistic towards others (low agreeableness), engage in antisocial behavior (low conscientiousness), and have unconventional interests (high openness). Notice that, since the possible combinations of the five personality traits are $3^{5} = 243$, the strong prevalence ($>10\%$) of a specific personality model in conflicting echo chambers is a very significant result.

\begin{table}[ht]
	\centering
	\begin{tabular}{c|cc|cc}
		& \multicolumn{2}{|c|}{\textbf{Science}} & \multicolumn{2}{|c}{\textbf{Conspiracy}} \\
		rank & PM & $\%$ & PM & $\%$ \\ 
		\hline
		1 & nynny & 14.57 & nynny & 17.66 \\ 
		2 & ooooo & 11.99 & nyony & 6.95 \\ 
		3 & nnnny & 5.69 & ooooo & 5.48 \\ 
		4 & oonny & 5.01 & nonny &  3.37 \\ 
		5 & nnony & 4.53 & oonny &  2.52 \\ 
		6 & nonny & 3.90 & nyyny & 2.41 \\ 
		7 & nyony & 3.58 & nnnny &  2.28 \\ 
		8 & onyoo & 3.39 & oynny &  2.24 \\ 
		9 & onnno & 2.58 & ynyon &  2.04 \\ 
		10 & nnyny & 1.96 & nnony &  2.01 \\ 
		\hline
	\end{tabular}
	\caption{\textbf{Prevalent Personality Models}. A personality model is characterized by five labels --- one for each of the \emph{Big Five} dimensions, i.e. extraversion, emotional stability, agreeableness, conscientiousness, openness --- indicating whether a user has a given personality trait (``y") or its reversed (``n") or none of the two (``o"). For instance, the personality model ``nyyoo" depicts users that are introvert, emotionally stable, and agreeable.}
	\label{tab:model}
\end{table}

Finally, we want to assess whether there is a correlation between users' activity and the emergence of certain personality traits. In both echo chambers, we observe very weak Pearson's correlations between the number of comments made by users and their personality traits.

\begin{table}[ht]
	\centering
	\begin{tabular}{ccccc | ccccc}
	 \multicolumn{5}{c|}{\textbf{Science}} & \multicolumn{5}{|c}{\textbf{Conspiracy}} \\
		E & S & A & C & O & E & S & A & C & O \\ 
		\hline
		 $-0.07$  & $0.06$ &  $-0.06$   & $-0.07$ &  $ 0.08$  & $-0.04$    & $0.06$  & $-0.04$  & $-0.04$ &  $0.06$ \\ 

	\end{tabular}
	\caption{\textbf{Correlation Analysis}. Pearson's correlations between the number of comments made by users and their personality traits i.e. extraversion (E), emotional stability (S), agreeableness (A), conscientiousness (C), openness (O) --- appear very weak in both the observed echo chambers. }
	\label{tab:2}
\end{table}

Such a result provides meaningful insights towards the relationship between the psychological profile of a user and his commitment inside a polarized online community. Indeed, the weak correlations between users' activity and their personality traits indicate that the permanence within echo chambers slightly shapes users' psychological profiles. Rather, our analysis suggests that the presence of specific personality traits in individuals lead to their considerable involvement in supporting narratives inside virtual echo chambers.

\section{Conclusions}
In online social media, users consume different information according to their preferences. Being influenced by confirmation bias and selective exposure, they join virtual polarized communities wherein they reinforce their preexisting beliefs.

In this paper, using a quantitative analysis on a massive dataset (more than $3M$ comments), we compare personality traits --- i.e. extraversion, emotional stability, agreeableness, conscientiousness, and openness --- of about $30K$ users embedded in different and conflicting echo chambers. 

Our results show that such personality traits are similarly distributed within the polarized communities, with the exception of the emotional stability, which is higher in users supporting the conspiracy-like narrative. Moreover, we find very similar and significant correlations between personality traits of users supporting conflicting narratives. Further, we show that the prevalent personality model is the same in both the echo chambers. In particular, the most common supporters of Science and Conspiracy tend to enjoy interactions with close friends (low extraversion), are emotionally stable (high emotional stability), are suspicious and antagonistic towards others (low agreeableness), engage in antisocial behavior (low conscientiousness), and have unconventional interests (high openness). Finally, we observe very weak Pearson's correlations between the number of comments made by users and their personality traits. Such a result suggests that the presence of specific personality traits in individuals lead to their considerable involvement in supporting narratives inside virtual echo chambers.

\section*{Acknowledgements}
The author is grateful to Fabiana Zollo, Michela Del Vicario, Antonio Scala, and Walter Quattrociocchi for valuable discussions. Moreover, special thanks to Geoff Hall and Skepti Forum for providing fundamental support in defining the atlas of Facebook pages disseminating conspiracy theories and myth narratives.

\bibliographystyle{unsrt}
\bibliography{biblio}

\end{document}